\documentclass[twocolumn,prl,aps]{revtex4}
\usepackage{amsmath}
\usepackage{amssymb}
\usepackage{graphicx}
\usepackage{dcolumn}
\usepackage{bm}
\usepackage{xcolor}
\usepackage{epstopdf}

\begin{document}
\title{Quantum transport of Dirac fermions  in  HgTe gapless quantum wells}

\author{G. M. Gusev, $^1$  A. D. Levin, $^{1}$ D. A. Kozlov$^{2}$  Z. D. Kvon, $^{2,3}$  and  N. N. Mikhailov $^{2,3}$}

\affiliation{$^1$Instituto de F\'{\i}sica da Universidade de S\~ao
Paulo, 135960-170, S\~ao Paulo, SP, Brazil}
\affiliation{$^2$Institute of Semiconductor Physics, Novosibirsk
630090, Russia}
\affiliation{$^3$Novosibirsk State University, Novosibirsk 630090,
Russia}

\date{\today}
\begin{abstract}

We study transport properties of HgTe quantum wells with critical well thickness, where the band gap is closed,
and the low energy spectrum is described by a single Dirac cone. In this work, we examined both macroscopic and micron-sized (mesocopic) samples.
In micron-sized samples, we observe a magnetic field induced, quantized resistance ($\sim h/2e^{2}$) at Landau filling factor $\nu=0$,
corresponding to the formation of helical edge states centered at the charge neutrality point (CNP). In macroscopic samples,  the resistance near zero Landau level (LL)
reveals strong oscillations, which we attribute to scattering between the edge $\nu=0$ state and bulk $\nu\neq 0$ hole LL.
 We provide a model taking an empirical approach to construct a LL  diagram based on a reservoir scenario, formed  by the heavy holes.

\end{abstract}

\maketitle
\section{Introduction}

The gapless helical edge states flowing along the edge of the two-dimensional (2D) system attract the attention of many due to
both fundamental and practical motivations. First, their existence serves as a signature for 2D topological insulators
\cite{bernevig, kane, hasan, qi, moore, moore2}. Second, the one dimensional nature of the edge states offers the possibility
to study strongly correlated fermion systems  such as the helical Tomonaga–Luttinger liquid \cite{kainaris}. Moreover, the helical edge states
 can be used to produce Majorana or parafermion modes for quantum computation  \cite{moore3}.

Helical edge states arise at the edges of the topological insulator (spin Hall effect) in the absence of an external magnetic field.
Particulary, the HgTe based quantum well with inverted band spectrum can host topological helical states \cite{zhou, konig, roth, gusev, gusev2}.
It is expected that these helical channels lead to quantized conductance with the value of $e^{2}/2h$ and nonlocal edge transport \cite{roth, gusev2}, which has been
observed only  for short distances between the measurement probes in the range of the few micrometers. The deviation between the theoretical prediction
and experimental value has been attributed to many different effects, including effects of Rashba spin-orbit coupling \cite{strom, crepin}, charge puddles \cite{vayrynen,vayrynen2}
and other numerous sources of inelastic scattering \cite{loss}.

Besides the insulator with a bulk gap, helical states can also exist in a gapless system.
 Such a remarkable example is 2D  massless Dirac fermions in the presence of a strong perpendicular magnetic field, such as graphene \cite{sarma, geim} and
 gapless HgTe quantum wells \cite{buttner, kozlov, kozlov2, gusev3, kozlov3}. It has been demonstrated that at the critical HgTe well thickness $ d_{c}$ equal to,
 depending on the surface orientation and the quantum well deformation, $6.3 - 6.5 $ nm, the band gap collapses, and  single-valley Dirac fermions
can be realized. In the presence of a strong perpendicular magnetic field, the zero Landau level of the Dirac fermions forms two counter propagating edge states similar to 2D topological insulators \cite{gusev3}. As a result, conductance is zero in the QHE regime and quantized in universal units $e^{2}/2h$  in the quantum Hall (QH)-metal regime in the absence of
backscattering between spin-polarized states.

In the present work, we studied the quantum transport in both  mesocopic  and macroscopic devices fabricated from HgTe zero-gap quantum structures.
In the mesoscopic samples, we observe a magnetic field induced, quantized resistance at $\nu=0$. These experiments clearly demonstrate the existence of a robust helical edge state in a system with single-valley Dirac cone materials.
In macrosopic sample, the resistance strongly deviates from the quantized value. Moreover, we observe large oscilations of the resistance at $\nu=0$.
We attribute these oscillations to the elastic intersubband scattering between the edge $\nu=0$ state and bulk $\nu\neq 0$ hole LL.
We observed a unconventional  LL  diagram for hole Dirac particles with several ring-like patterns, which is attributed
to LL crossing of single LL and manifold-degenerate subband levels.  We report a model taking into account the reservoir
of the sideband hole states. The model reproduces some of the key features of the data, in particular the density  dependence
of the  hole LL and manifold LL crossing points. We believe that this model provides a framework for more sophisticated theoretical
tools to understand many-body phenomena, such as spin-splitting enhancement effects.

\section{Materials and Methods.}

Quantum wells $Cd_{0.65}Hg_{0.35}Te/HgTe/Cd_{0.65}Hg_{0.35}Te$ with (013) surface orientations and a well thickness of 6.3-6.4 nm
were prepared by molecular beam epitaxy.
Two different types of devices were used: macroscopic and mesoscopic Hall bars.
The mesoscopic sample is a Hall bar device with 2 current and 7 voltage probes. The bar has a width $W$ of $3,2 \mu m$
and three consecutive segments of different lengths $L$ $(2.8, 8.6, 33 \mu m )$.
The macroscopic bar has a width W of $50 \mu m$ and three consecutive segments of different lengths L(100, 250, 100$\mu m$).
A dielectric layer was deposited (100 nm of $SiO_{2}$ and 100 nm of $Si_{3}Ni_{4}$) on the sample surface and
then covered by a TiAu gate. The density variation with gate voltage was $1\times10^{11} cm^{-2}V^{-1}$. The magnetotransport measurements were performed in the temperature range
$1.4-4.2 K$  using a standard four point circuit with a $1-27 Hz$ ac current of $1-10 nA$ through the sample,
which is sufficiently low to avoid overheating effects.  6 devices from the different wafers were measured, all with similar results.

\begin{figure}[ht]
\includegraphics[width=\linewidth]{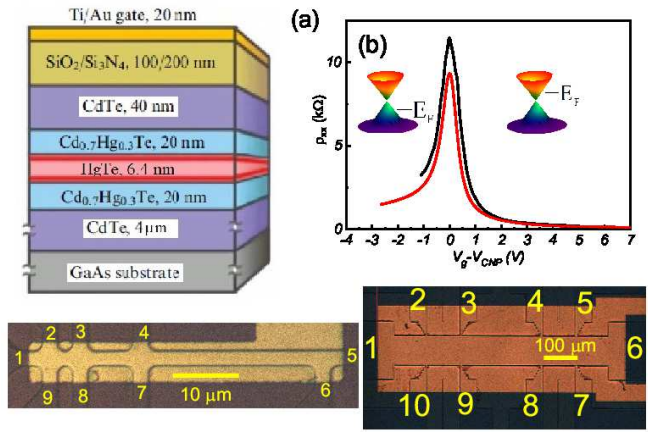}
\caption{\label{fig.1}(Color online) (a) Schematic of the transistor. (b) Resistivity $\rho_{xx}$ as a function of gate voltage measured for different devices.
The red trace-macroscopic, black line- mesocopic devices.
The bottom of the figure presents a top view of the samples.}
\end{figure}

\section{Transport measurements in micron-sized (mesoscopic) samples.}

The variation of resistivity with gate voltage and lattice  temperature for 6.4 nm  quantum wells for mesoscopic and macroscopic devices is shown in Figure 1b.
The current flows between contacts 1,5; voltage is measured between probes 2,3 ($\rho_{xx} = \frac{W}{L}R_{xx}, R_{xx}=R^{2,3}_{1,5}=V_{2,3}/I_{1,5}$) for mesoscopic device;
and current is applied between contacts 1,6; voltage is measured between probes 2,3 $R_{xx}=R^{2,3}_{1,6}=V_{2,3}/I_{1,6}$ for macroscopic device.
The resistance behaviour in zero magnetic field resembles behaviour in other HgTe-based quantum wells, including topological insulators \cite{konig, gusev, gusev2}:
resistance shows a peak around the charge neutrality point (CNP). In graphene and zero gap HgTe wells, the CNP is coincident with the Dirac point.
Transport in HgTe QWs of a critical width $d_{c}$ is expected to be determined by the energy gap fluctuations
leading to the formation of the topological channel network \cite{entin,gusev4}.
The minimum conductivity at the Dirac point $\sigma_{xx}=1/\rho_{xx}=\frac{e^{2}}{h}(2.5\pm 1)$ agrees with the observations \cite{gusev4}.

In the presence of a magnetic field, the energy spectrum is organized in Landau levels (LLs)
with square root versus linear dependence on the magnetic field and square root dependence on the Landau index $n$.
Moreover, there is an additional zero energy LL, originated from Berry phase carried by each Dirac point, similar to graphene  \cite{castro}.
It is worth noting that, in HgTe quantum wells, Dirac fermions have a single cone (one valley) spectrum,
which allows the realization of edge state transport in a strong magnetic field via counter propagating modes \cite{gusev2},
while in graphene transport depends on which degeneracy, spin or valley, is removed first in a strong magnetic field \cite{abanin}.
A symmetric  LL spectrum around zero energy level is expected for low energy.

\begin{figure}[ht!]
\includegraphics[width=\linewidth]{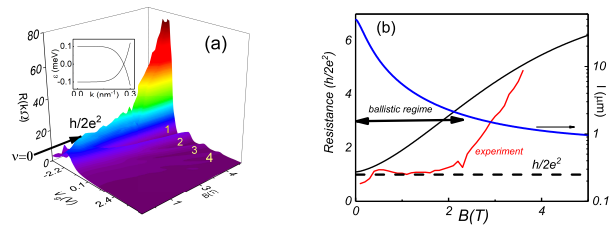}
\caption{\label{fig.2}(Color online) The color map of $R_{xx}(N_{s},B)$ versus $N_{s}$ and B at T=4.2K for mesoscopic device. The arrow indicates the plateau of resistance at $\nu=0$.
The insert shows the  counterpropagating spin-polarized edge states in the presence of a strong perpendicular magnetic field.
 (b)The red trace represents the longitudinal $R_{xx}$  resistance as a function of the magnetic field (B) at the CNP. The black line shows
the theoretical resistance calculated from the model \cite{calvo}. The blue line represents the B dependence of the mean free path calculated from the model \cite{calvo}. }
\end{figure}

In this section we present the results for the mesoscopic sample.
Longitudinal $R_{xx}$  resistance has been measured as a function of gate voltage ($V_{g}$)  and magnetic field (B).
Figure 2 shows the the resistance  color plots as a function of carrier density and B. One can see stripes corresponding to resistance maxima and minima in the B,
and the slopes of the stripes are determined by the LL filling factor $\nu$ : $dN_{s}/dB=\nu e/h$, where $h$ is the Plank constant.
The Dirac point corresponds to the charge neutrality point (CNP), where Hall resistance passes zero value and changes sign \cite{gusev2}.
The zero energy Landau level occurs at CNP splitting, due to Zeeman energy at the high magnetic field, which leads to the formation of two counter propagating states (insert in Figure 2a).
Simultaneous observations of the resistance plateaux in local and in nonlocal (not shown) transport confirms this
scenario \cite{gusev2}.  The  experimental consequences expected for the ballistic edge transport resulting from the helical states  is resistance quantization with the universal value $\frac{h}{2e^{2}}$.
Figure 2b shows the resistance trace corresponding to the chemical potential position at the CNP. One can see that the resistance
plateau reaches the quantized value   $\frac{h}{2e^{2}}$ in the range of the magnetic field $0.5 < B < 2 T$, diverging towards the insulating value at higher B.
The resistance quantization is not perfect and demonstrates mesoscopic fluctuations similar to resistance fluctuations observed in
2D topological insulators in zero magnetic field \cite{konig, gusev}. As was mentioned above, the mechanism of resistance deviations in TI
is still under discussion \cite{loss}. Note, however, that in the presence of a strong magnetic field and spin orbit interaction,
backscattering between different spin polarized chiral edge channels may occur \cite{khaetskii}. Adapting this model for the
helical edge states and assuming scattering by the Coulomb impurities  in the presence of the spin orbit coupling, we can obtain
the equation for inverse scattering length \cite{calvo}
$$
\frac{1}{l}=\frac{(2\pi)^{3/2}}{v_{1}v_{2}}\left[\frac{e^{2}}{\hbar\varepsilon}\right]^{2}\frac{N}{q_{s}^{2}\lambda}\left[\frac{m\delta v \alpha g \mu H}{\delta E^{2}}\right]^{2}
$$

where $v_{1,2}$ are the velocities of the spin polarized edge states,  $\delta v= v_{2}- v_{1}$, N is the impurity density, $q_{s}$ is the inverse screening length,
$\lambda=\sqrt{\hbar c/eB}$ is the magnetic length, $g \mu H$ is the Zeeman term. The energy splitting between edge states
is determined by equation
$$
\delta E=\sqrt{(g \mu H/2)^{2}+ (mv\alpha)^{2}}
$$
where $v$ is the averaged edge state velocity, $\alpha$ is spin orbit coupling constant, $m$ is the effective mass. Assuming $\delta v \approx \delta E/\hbar \omega _c \ll v$,
where $ \omega _c =eB/mc$ is the cyclotron frequency, we calculate the scattering length for our system. Figure 2b demonstrates the
magnetic field dependence  of the scattering length for parameters: $N=10^{11} cm^{-2}, \alpha\approx 10^{5} m/s, v=10^{5} cm/s$. The resistance can be calculated from equation
$R=h/2e^{2}(1+L/l)$ \cite{abanin, gusev}. One can see that the characteristic scattering length strongly decreases with the magnetic field and
becomes comparable with the distance between probes at $B_{c}\approx2.5 T$. Therefore, the transport regime is expected to be ballistic
below $B_{c}$, and resistance is quantized, while  the resistance increases at $B>B_{c}$  because the electrons experience more scattering.

\section{Transport measurements in macrosopic samples.}

In this section, we focus on the transport properties in large-sized macroscopic samples.
Figure 3a shows the
the resistance (a) color plots as a function of carrier density and B for a smaller field range. In addition we invert the gate voltage
 scale in comparison with Fig.2a and demonstrate the hole-like spectrum of the LL on the right side of the voltage sweep.
 One can see a significant difference between microscopic and macroscopic sample behaviour near the CNP: the resistance in the small sample
 is quantized and shows the plateau at low field , while the resistance in the large device is much larger than the value $h/2e^{2}$
 and reveals oscillations. Fig. 3b shows the evolution of resistance at the CNP with the magnetic field. Note, that at the CNP Hall resistance
 is zero, therefore,  the behaviour of the transport coefficients in the quantum Hall effect regime and at $\nu=0$ are very different.
 For example, when $\rho_{xy}\gg \rho_{xx}$ one can expect that $\rho_{xx}\sim \sigma_{xx}$ . In contrast at  $\nu=0$ we observe  $\rho_{xy}\approx 0$,
 and $\rho_{xx}\sim 1/\sigma_{xx}$. In Fig. 3b we plot conductivity versus the magnetic field, and, for comparison, we also plot
 the B-dependence of resistivity at $V_{g}=-2.5 V$, corresponding to the quantum Hall regime of hole-like Landau levels. One can see
 the coincidence between the position of the conductivity peaks at $\nu=0$ and the resistivity (or conductivity, because $\rho_{xx}\sim \sigma_{xx}$ ) peaks of 2D Dirac-like holes.

To get more insight into the physics of the observed resistivity oscillations, it is important to consider the energy spectrum of of the gapless HgTe quantum well.
The particle energy in 6.4 nm HgTe wells represents a single valley cone and, aside from
 the Dirac-like holes in the center of the Brillion zone, the valance band contains local a valley formed by the heavy holes with a parabolic spectrum.
Therefore, one can expect that LL energy in the presence of the magnetic field is asymmetric for electrons and holes \cite{buttner,kozlov2, gusev3, kozlov3}.
A previous study of the quantum Hall effect in HgTe wells demonstrated strong asymmetry between electrons and holes,
which was attributed to the presence of a band maximum in the spectrum of the holes \cite {kozlov}.
It has been found that the quantized Hall plateaux for hole-like particles occurs in magnetic field 3 times smaller
than for electron-like carriers, and the plateaux for holes is much wider that for electrons.
The authors attributed such anomalous behaviour to the existence of sideband holes, which may serve as reservoir and pin the Fermi level in the gap between the Landau
levels of the Dirac holes. Recently heavy hole density of the states has been measured by the capacitance technique \cite{kuntsevich}.

Figure 4a shows a two-dimensional color plot of the hole part of the spectrum.
One can see, that the resistance  reveals a strikingly rich ring-like structure. Instead of  stripes, expected in a conventional LL diagram,
 sharp abrupt bends occur at low magnetic field. The empirical linear fit is shown as dashed lines in Fig.3a and basically corresponds to the
Dirac-like hole LL. The slopes of the lines decrease with the magnetic field, however,
it is always larger than $\nu e/h=2.4\times10^{10} cm^{-2}/T$. For example, one can see that the slope of the first LL
is close to $40\times10^{10} cm^{-2}/T$, which is about 17 time larger than expected.

\begin{figure}[ht!]
\includegraphics[width=\linewidth]{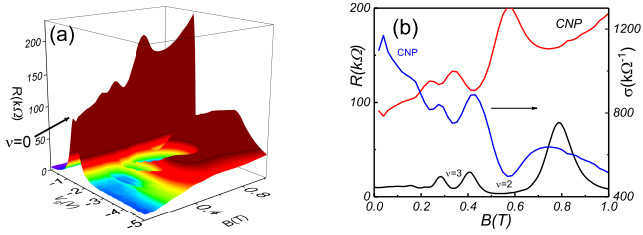}
\caption{\label{fig.3}(Color online) (a)  (a) Color map of $R_{xx}(N_{s},B)$ versus $N_{s}$ and B at T=4.2K for a macroscopic device.
(b) The red trace represents the longitudinal $R_{xx}$  resistance as a function the magnetic field (B) at the CNP.
The blue trace represents the conductivity $\sigma_{xx}$   as a function the magnetic field (B) at the CNP.
 The black trace represents $R_{xx}$ in the quantum Hall effect regime for 2D Dirac holes as a function the magnetic field (B) at $V_{g}=-2.5V.$  }
\end{figure}

\begin{figure}[ht!]
\includegraphics[width=\linewidth]{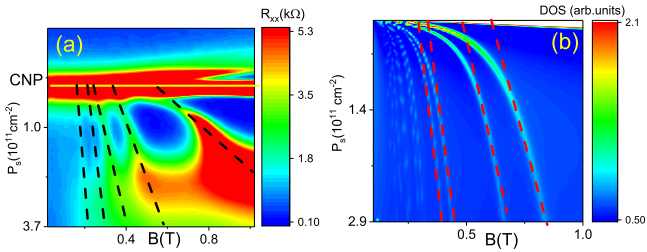}
\caption{\label{fig.5}(Color online) (a) Color map of $R_{xx}(N_{s},B)$ versus $N_{s}$ and B at T=4.2K for a macroscopic device. Black dashes represent
the slope of the LL for Dirac-like holes.
(b) Theoretical calculations of the density of states as a function of the hole density and magnetic field. Red dashes represent the slope of the LL for Dirac-like holes. }
\end{figure}

\begin{figure}[ht!]
\includegraphics[width=\linewidth]{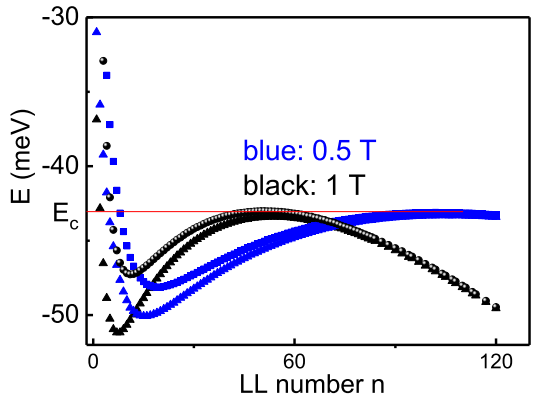}
\caption{\label{fig.6}(Color online) Calculated Landau levels for a 6.4 nm symmetric HgTe quantum well for B=0.5T and B=1T.
Two sets of levels originating from spin splitting of 2D subbands are shown. Horizontal lines show the energy when the
fermi level is pinned by the backside hole LLs. }
\end{figure}

It is worth noting that this unconventional LL pattern has never been  observed before in other 2D systems. Indeed,
in the presence of two subbands, the LL diagram shows a ring-like structure due to LL crossing \cite{zhang} with topology, which is different from our observations.
The LL crossing points become crossing 2-folds owing to the crossing between spin-split first and second subbands.
Instead of a diamond structure, expected from a naive picture, nonmonotonic behaviour of electrochemical potential
leads to a ring-like shape, although the electron-electron interactions may play a significant role too.
The LL spectrum in the valence band of a HgTe well becomes complicated at high energy and LL crossing occurs.
We use all relevant Kane Hamiltonian parameters to numerically calculate the density of the states  for the valence band
as a function of $P_{s}$ and $B$. The magnetic field and density coordinates of the LL crossing in the plot
 of the density of the states
correspond to the energy level crossing, and comparison with the experiment allows to determine
the Kane Hamiltonian parameters. However, one can see that the  calculations  show the LL crossing
at high B and density, and therefore, it is clearly insufficient to explain the ring-like pattern at low B and $P_{s}$
obtained in experiment.

 The features observed in our experiment can be understood  from the consideration of the behaviour of the chemical potential $\mu$
(the Fermi energy at T=0) in the presence of the reservoir formed by the density of the states originated from the tails of high index valence band LLs.
 To account for the key features of the model, it is important to get an idea of how the energy spectrum is quantized in the magnetic field.
Fig.5 shows the calculated Landau level originated from the valence band  for two fixed magnetic fields.
The low index levels  rapidly go up with increasing B.  In contrast, the high index levels form a dense set,
 especially near the band extreme, and are slowly shifted with increasing magnetic field. One can see that the number
 of levels near the maximum inside of the energy interval $\Delta E \approx5 meV$ is close to 80 at B=0.5 T.
 As the magnetic field increases further, the level number near the maximum decreases.
 The behaviour of the Fermi level in a two-dimensional system strongly depends on the density of the states. In conventional 2D
 electron gas, the Fermi level is proportional to the charge concentration because the density
 of  the states is constant, while in the system with a linear Dirac-like spectrum, $E_{F}$ is proportional
 to the square root of $N_{s}$. In the magnetic field,  $E_{F}$ jumps  from one Level to the next lower level. Deep
 minima in the diagonal resistance accompanied by a plateaux in $R_{xy}$ are attributed to the existence of localized
 electronic states on the tail of the broadened LL in the presence of the disorder and pinning of the Fermi level.
 Due to the big density of LL, shown in Figure 4, the Fermi level becomes locked near $E_{c}\approx -15 meV$ (the energy at the CNP corresponds to E=-30 meV),
 indicated by the red line, and a further increase of density results in the overlapp between heavy holes and Dirac-like holes.
 Note that the energy $E_{c}$ corresponds to the very low density $P_{s}=0.15\times 10^{11} cm^{-2}$.

 To calculate the color map plot of the density of the states $D$ as a function of density and B, we adapted the Lorentzian
 form of the density of the states in a strong magnetic field \cite{duarte}. Figure 4b shows the color map $D(N_{s}, B)$, assuming
 level broadening independent of the magnetic field. One can see two parts of the spectrum: the low density part consists of the
 stripes with the large slope, corresponding to the Dirac-like hole LL at $\mu < E_{c}$, and the high density part corresponds
 to the overlap between Dirak-like and heavy holes with the parabolic spectrum at $\mu > E_{c}$. We also plot the slope of the LL for Dirac-like holes at high densities,
 corresponding to the region where $\mu > E_{c}$.

 Now we turn to a detailed comparison between the experimental resistance plot of $R_{xx} (P_{s}, B)$ and the theoretical DOS for the LL
spectrum. In the experimental fan chart, we don't see the slope for the Dirac-like hole Landau levels at $\mu < E_{c}$ because of the broadening
of the zero LL. We can resolve the LL only in a very narrow energy (density) window $0.2\times10^{11} < P_{s} < 0.2\times10^{11}$.
For higher densities, LL broadening abruptly increases and particle motion becomes not quantized into discrete levels. Our model
is much too simple to adequately describe the slope for the Dirac-like hole Landau levels for all densities and more advanced theory is required to
describe this behavior, which is out of the scope of our experimental paper. However the model can qualitatively explain the difference
of the LL slope from the one expected.

As we demonstrated above, conductivity at the CNP reveals the oscillations, which coincide with conductivity oscillations close in proximity to the CNP (Fig.3b).
We attribute this effect to the resonance scattering between helical edge states at $\nu=0$ to the bulk LL. In such narrow band gap materials as HgTe,
potential fluctuations due to nonuniform doping play a significant role. Such potential fluctuations  lead to the formation of conducting large size puddles or lakes in the
bulk of the insulator, and carriers at the edge states interact with these puddles \cite{vayrynen, gusev2}. A number of puddles should be present in the vicinity of
the edge to allow for scattering between the counter-propagating states and the bulk LL localized in each lake.

\section{Conclusions}

In this paper, we present a detailed study of the transport in single cone Dirac fermions in 6.3-6.4 nm HgTe quantum wells in mesoscopic and
macroscopic devices. We observe quantized four-terminal resistance in mesoscopic devices which  provides a  stark indicator for helical edge transport at $\nu=0$
in the presence of a magnetic field.
In macroscopic samples, we observe resistance oscillations at $\nu=0$ and  an unconventional  LL  diagram for hole Dirac particles with several ring-like patterns. We  attribute
the fan chart to LL crossing of single LL and manifold-degenerate subband levels.  We report a model taking into account the reservoir
of the sideband hole states. The model reproduces some of the key features of the data, in particular the density  dependence
of the hole LL and manifold LL crossing points. The oscillations of resistance at the CNP may occur due to the elastic intersubband scattering between the edge $\nu=0$ state and bulk $\nu\neq 0$ hole LL
localized in the large size puddles near the edge.

\section{Acknowledgment}
We  thank  O.E. Raichev  for  the  helpful  discussions and providing us with some of the calculations of the LL.

\end{document}